# Security Issues in Data Warehouse

SAIQA ALEEM      LUIZ FERNANDO CAPRETZ

Department of Electrical & Computer Engineering
Western University
London, ON, Canada
{saleem4, lcapretz}@uwo.ca

FAHEEM AHMED

Department of Computing Science
Thompson Rivers University
Kamloops, BC, Canada
fahmed@tru.ca

*Abstract* — Data Warehouse (DWH) provides storage for huge amounts of historical data from heterogeneous operational sources in the form of multidimensional views, thus supplying sensitive and useful information which help decision-makers to improve the organization's business processes. A data warehouse environment must ensure that data collected and stored in one big repository are not vulnerable. A review of security approaches specifically for data warehouse environment and issues concerning each type of security approach have been provided in this paper.

## 1 Introduction

In today's competitive business environment, organizations need to collaborate with each other and track their performance for market trend analysis. With the help of advances in computer and network technology, organizations stores, collect, and analyze vast amounts of data efficiently and quickly. Data are analyzed by the organization not only for market trend identification, but also to examine the effectiveness of their activities and to make decisions that affect their bottom line. Therefore, data management has become crucial because organizations not only need to store and retrieve data, but also need to derive meaningful information from it. As a consequence, organizations have come to depend more on knowledge management technologies such as interoperable knowledge management, knowledge repositories, and data warehouses (DWH).

A data warehouse may contain massive amounts of organizational data such as financial information, credit card numbers, organization trade secrets, and personal data, thus they are vulnerable to cyber attack [1]. A DWH must ensure that sensitive data does not fall into the wrong hands when data are consolidated into one big repository and become an easy target for malicious outside or inside attackers. Many published security statistics show that the number of attacks on data is increasing continuously [2]. Data security focusses mainly on three issues: confidentiality, integrity, and availability, these concepts are also know by the acronym CIA.

Confidentiality emphasizes protection of information from unauthorized disclosure, either by indirect logical inference or by direct retrieval [3]. Integrity involves data protection from accidental or malicious changes such as false data insertion, contamination, or destruction. Availability ensures that data are accessible to all authorized users at any time. In the past, many data security solutions for databases have been proposed.

Although available solutions have been proven to be scientifically effective, they are infeasible or at least inefficient for a DWH environment because this environment requires specific performance. Most of today's DWH security solutions lack effective security procedures to protect the data accessed through them. Existing security methods can be best for restricting security breaches, but cannot completely eliminate the risk.

In this paper, we present a survey of the security approaches available for DWHs and the issues concerning each type of security approach. The remainder of the paper is organized into two sections. In Section 2, various existing data security solutions for DWHs are presented, and specific issues in the DWH environment are discussed. Finally, Section 3 concludes the paper and highlights future research directions.

## 2. Security Approaches for DWH

A DWH is an integral part of an organization and empowers its users by enabling them to retrieve information about the business process as a whole. According to Devbandu [4], security is an important requirement for DWH development, starting from requirements and continuing through implementation and maintenance. Security solutions for on-line transactional processing (OLTP) systems cannot be appropriate for DWHs because in OLTP, security controls are applied on rows, columns, or tables, while DWHs need to be accessed by different numbers of users for different content because multidimensionality is a basic principle of a DWH [1, 5].

Data extraction, transformation, cleaning, and preparation have all been done before the data are loaded into the DWH. Security concerns must be addressed at all layers of a DWH system. Moreover, DWH security cannot be ensured unless the security of the underlying operating system and the network have been addressed [6]. Various security solutions have been proposed in the DWH literature and are described below, categorized according to how they address basic security concerns such as CIA.

## 2.1 DWH Security Approaches for Confidentiality Issues

Confidentiality emphasizes protection of information from unauthorized disclosure, either by indirect logical inference or by direct retrieval [3]. In order to address DWH confidentiality concerns, many approaches have been proposed dealing with access control. Access-control mechanisms involve controlling both invocation and administration of the DWH and the source databases. Authentication and audit mechanisms also fall under access control and must be installed in a DWH environment.

Conventionally, DWHs have been accessed by high-level users such as business analysts and executive management. Therefore, critical access-control issues also arise at the front end of a DWH. Most DWH or OLAP vendors assume that there is no need to provide fine-grained access-control support for a DWH front end because it hinders discovery of analytical information. However, this assumption is not appropriate because many users can access analytical tools to query the DWH. Front-end DWH applications can provide both static and dynamic reporting. Imposing access control on static reports is not a problem because it can be defined on a report basis. For dynamic reporting like data-mining queries, it is difficult to provide appropriate access-control policies. This leads to the problem of data inference; for example, a user may not be authorized to obtain particular information, but may retrieve it through an aggregated query.

## 2.2 DWH Security Approaches for Integrity

Integrity involves data protection from accidental or malicious changes such as false data insertion, contamination, or destruction. The disadvantage of access-control mechanisms is that they do not capture inferences on data in the case of an aggregated OLAP query. Inferences on data lead to the integrity issue. For more than thirty years, inference-control approaches have been studied in statistical and census databases [7, 8, 9]. The proposed approaches can be categorized into restriction-based and perturbation-based techniques. Restriction-based inference control techniques simply deny unsafe queries to prevent malicious inference. Perturbation techniques add noise to data, swap data, or modify the original data and can also apply data modification to each query dynamically. The approaches presented to solve the integrity issue can be classified further as described below.

### 2.2.1 Restriction-based approaches

In restriction-based inference-control techniques, the safety of a query is determined based on the maximum number of values aggregated by dissimilar queries [8], the minimum number of values aggregated by a query [10], and the highest rank of the matrix expressing answered queries [11].

Micro-aggregation and partitioning considers specific type of aggregations. In partitioning methods, a partition is defined on sensitive data, and a restriction is applied on a complete block of a partition for aggregate queries [12, 13]. Micro-aggregation also replaces cluster averages with their sensitive values [14]. Both methods are not based on dimensional hierarchies and therefore may contain meaningless blocks that are not useful for users.

### 2.2.2 Combined Access- and Inference-Control Approaches

In order to remove security threats, access control and inference control together can provide a good solution. Ensuring security should not affect the usefulness of DWH and OLAP systems. Wand & Jajodia [15] proposed a three-tier security architecture for a DWH. Usually, two tiers can be found in statistical databases, such as sensitive data and aggregation queries. This two-tier architecture has some inherent drawbacks: inference checking during run-time query processing may result in unacceptable delays, and also under this two-tier architecture, inference-control techniques cannot benefit from the special characteristics of OLAP. To overcome these drawbacks, the research has defined a three-tier architecture to provide access control between the first and second tiers and inference control between the second and third tiers.

The basic lattice-based inference method [16] can be used and implemented on the three-tier inference-control model. The first methodology used existing inference-control methods for statistical databases, whereas the second methodology was designed to remove the limitations of existing inference-control methods. The work claims that both methods could be applied on the basis of a three-tier inference control architecture that is more appropriate for DWH and OLAP systems specifically.

### 2.2.3 Modelling-based Approaches to DWH Security

Triki *et al.* [17] proposed approach provides semi-automatic inference detection at the DWH design level. The approach presented consists of three phases. The first phase identifies sensitive data from DWH schemata with the collaboration of security designers and experts in the field. In the second phase, an inference graph based on a class diagram is constructed to detect elements which may cause inferences in future. The security designer also distinguishes between elements leading to precise and partial inferences. Precise inference means that exact information is disclosed, whereas partial inference leads only to partial disclosure of information.

The inference graph consists of a set of nodes representing the data. Then nodes are connected to each other by oriented arcs representing the direction of inference and its type (partial or precise). In the third phase, DWH schemata are enriched automatically by UML annotations which flag the elements that may lead to both types of inferences. The work claimed that their approach had two advantages: independence of the data domain, and use of available data to detect inferences.

### 2.2.4 Data Masking and Perturbation-Based Security Approaches

Data disclosure can be easily avoided by data-masking approaches. Using data masking, original data values can be replaced or changed. Currently, the best practices for data masking are used by Oracle in their DBMS [18]. In data masking, encryption is an advanced form of enforcing privacy. Oracle has also developed Transparent Data Encryption (TDE) in the 10g and 11g versions of its DBMS. TDE incorporates the well-known AES and 3DES encryption algorithms [19, 20].

Santos *et al.* [21] proposed a data-masking technique for data warehouses consisting only of numerical values. The proposed approach was based on mathematical modulus operators such as division, remainder, and two simple arithmetic operations, which can be used without changing DBMS source code and user applications. They claimed that the proposed formula required low computational effort and that as a result, query response-time overheads became relatively small while still providing an appropriate security level.

## 2.3 DWH Security Approaches for the Availability Issues

Data availability is of utmost importance in any DWH system. This involves data recovery from real-time corruption or incorrect data modification and continuous 24/7 user access. Data replication is performed to be able to restore damaged data using many proposed solutions. In this way, database downtime because of maintenance interventions can also be avoided, and query-processing efforts can be divided, avoiding data-access hotspots. Well-known RAID architectures can be used for mirroring data [22, 23] on systems where centralized servers contain the database. However, organizations have been implementing their DWHs in low-cost machines for cost-optimization purposes. RAID technology is not suitable for this kind of situation because typically only one disk drive is present.

In today's market, commercial solutions for the DWH data-availability issue are available, such as Oracle RAC [24] and Aster Data [25]. Hamming codes provide another approach to recover corrupted data using error-correction codes. The proposed data-storage system makes it possible to recover corrupted data blocks by using error-correcting codes, remapping bad blocks, and replicating blocks [26, 27]. Marsh & Schneider [28] proposed a technique for distributed storage used the same features as described earlier plus encryption methods. Other researchers [29, 30, 31, 32, 33] have also proposed architecture assessment and self-healing methods to address the availability issue. Recently, Darwish *et al.* [34] have establish cloud-based protocols to defend against denial-of-services attacks.

## 3 Discussion

A literature review of the various approaches to DWH security has been presented above. A DWH needs powerful security features in addition to its normal functionalities. The primary security requirements are summarized by the Confidentiality, Integrity and Availability (CIA) acronym. A full set of security features can be defined under these three basic properties, such as access control, inference control, non-repudiation, authentication, authorization, and availability. The best security model is one that provides end-to-end security in all phases of DWH, starting from modelling and continuing through implementation and maintenance. Moreover, the security model must address the three basic CIA security requirements. Some of these approaches consider security requirement confidentiality.

Security approaches which discussed integrity issues were further classified by how they address this type of security concern. Some of the approaches also tried to address the issue of DWH availability. In short, all the proposed approaches addressed only some aspects of security, and a DWH security model are needed that covers all the security requirements and also help in developing a secure DWH. The identified issues with security approaches in DWHs are listed below:

a) Proper identification of security policies is a highly critical starting point in implementing security in a DWH.

b) Most of the approaches used standard encryption methods and tried to provide strong data privacy. However, use of this type of encryption method makes them inefficient for DWH use. Encryption algorithms like AES and 3DES require large computational effort and have a huge impact on performance. A technique is therefore needed that provides strong data privacy with less computational effort and also maintains high performance, which is the basic requirement of DWH use.

c) A method is also needed that specifically addresses the DWH availability issue. It will improve existing data-recovery methods to repair or restore corrupted data quickly, efficiently, and effectively.

d) Evaluation methods for DWH security are also needed. None of the approaches examined addresses the issue of how one can assess the maturity level of security in a DWH.

e) Confidentiality, data integrity, and availability are also basic requirements for DWH security. A combination of the approaches discussed above could be helpful in providing a solution to this problem.

f) Most of the approaches are domain-dependent, not generic, or are somehow constraints-based.

g) A DWH security maintenance mechanism is needed that takes specific security requirements into consideration and applies them appropriately.

h) A model is needed that helps to identify security requirements automatically throughout the DWH life cycle and makes it possible to provide proper authentication.
None of the existing approaches addressed this issue. The proper identification of security policies is a highly critical starting point in implementing security in a DWH.

i) Most of the approaches used standard encryption methods and tried to provide strong data privacy. However, use of this type of encryption method makes them inefficient for DWH use. Encryption algorithms like AES and 3DES require large computational effort and have a huge impact on performance. A technique is therefore needed that provides strong data privacy with less computational effort and also maintains high performance, which is the basic requirement of DWH use.

In order to provide DWH security, the real goal is to protect data Security and to preserve an appropriate level of privacy requirements must be considered in all layers of the system involved. No efforts have been made until now to integrate security into the complete DWH development cycle. Some approaches consider security requirements in the early stages of the DWH development life cycle. More efforts have been put in logical modelling of DWH security requirements, but they have not provided any tool support for implementing the modelled security requirements automatically in the target DWH system. A holistic approach of security throughout the software life cycle [35], may also benefit from a neuro-fuzzy framework [36, 37] - like it has been applied to other application domains.

## 4. Conclusion

This study has provided a literature review of existing DWH security solutions, discussing their issues and their impact on DWH scalability and performance requirements. It has become apparent that the proposed solutions are infeasible or inefficient for use in DWH environments. A DWH requires specific functionality with tight scalability and performance requirements. A complete solution is therefore needed that makes it possible to address these directives. DWH security is an active research relevance to any industrial project. Further research in DWH security is needed to address the issues discussed above because many more aspects remain to be considered, and there many open questions to be answered.